\documentclass[runningheads]{llncs}
\usepackage{graphicx}
\usepackage{amssymb}
\usepackage{xcolor}
\usepackage{pgf,tikz,pgfplots}
\usepackage{amsmath}
\usepackage{comment}  
\usepackage{multicol}
\usetikzlibrary{shapes,decorations,arrows,calc,arrows.meta,fit,positioning}
\usepackage[algo2e,linesnumbered,ruled,lined]{algorithm2e}

\newtheorem{assumption}{Assumption}

\newcommand{\Ker}{T_{\mathfrak s}}

\newcommand{\Sadegh}{\textcolor{magenta}}

\DeclareMathOperator{\Paths}{\mathsf{PATH}}

\newcommand{\Cont}{\mathcal{C}}

\usepackage[shortlabels]{enumitem}

\newcommand{\alphabeth}{\Sigma}

\newcommand{\AP}{\mathsf{AP}}            

\newcommand{\always}{\Box}
\newcommand{\eventually}{\Diamond}
\newcommand{\until}{\mathbin{\sf U}}
\newcommand{\release}{\mathbin{\sf R}}
\newcommand{\nex}{\mathord{\bigcirc}}

\newcommand{\word}{\boldsymbol{\omega}}
\newcommand{\wordt}[1]{{\omega}_{#1}}

\newcommand{\CMP}{\mathfrak{S}}
\newcommand{\StS}{\mathcal{S}} 
\newcommand{\IS}{\mathcal{U}} 

\newcommand{\hatS}{\hat{\mathcal{S}}} 
\newcommand{\Kerp}{T_{\mathfrak p}}
\newcommand{\Kero}{T_{\mathfrak o}}

\newcommand{\LDBA}{\mathcal{A}}
\newcommand{\Acc}{\textsf{Acc}}

\newcommand{\Lab}{\textsf{L}}   

\usepackage{mathtools}
\newcommand\myeq{\stackrel{\mathclap{\normalfont\mbox{($\ast$)}}}{=}}
\newcommand\mygeqq{\stackrel{\mathclap{\normalfont\mbox{($\ast\ast$)}}}{\ge}}

\begin{document}
\title{Formal Policy Synthesis for Continuous-Space Systems via Reinforcement Learning
}
\titlerunning{Formal Policy Synthesis for Continuous-Space Systems via RL}
%
\author{Milad Kazemi
\and
Sadegh Soudjani
}

\authorrunning{M. Kazemi and S. Soudjani}
\institute{School of Computing, Newcastle University\\
Newcastle upon Tyne, United Kingdom
}
%
\maketitle              
\begin{abstract}
This paper studies satisfaction of temporal properties on unknown stochastic processes that have continuous state spaces. We show how reinforcement learning (RL) can be applied for computing policies that are finite-memory and deterministic using only the paths of the stochastic process. We address properties expressed in linear temporal logic (LTL) and use their automaton representation to give a path-dependent reward function maximised via the RL algorithm. We develop the required assumptions and theories for the convergence of the learned policy to the optimal policy in the continuous state space. To improve the performance of the learning on the constructed sparse reward function, we propose a sequential learning procedure based on a sequence of labelling functions obtained from the positive normal form of the LTL specification. We use this procedure to guide the RL algorithm towards a policy that converges to an optimal policy under suitable assumptions on the process. We demonstrate the approach on a 4-dim cart-pole system and 6-dim boat driving problem.
\end{abstract}
%
%
%

\section{Introduction}
\label{sec:Intro}

\textbf{Motivations.}
Omega-regular languages provide a rich formalism to unambiguously express desired properties of the system. Linear temporal logic (LTL), as a class of omega-regular languages, is widely used for task specification such as safety, liveness, and repeated reachability. Synthesising policies formally for a system to satisfy a specification requires the knowledge of a model of the system. Extensive techniques are developed in the literature for different classes of models including finite-space models \cite{BK08} and continuous-state or hybrid models \cite{HS_TAC19,lavaei2020formal,majumdar2019symbolic,mallik2017compositional,Jagtap2019}.
Reinforcement learning (RL) is a promising paradigm for sequential decision making when a model of the system is not available or is very hard to construct and analyse. The objective of an RL algorithm is to find suitable action policies in order to maximise the collected rewards that depend on the states and actions taken at those states. The RL algorithms are in particular useful when the total collected reward has an additive structure. 

Many objectives including satisfaction of omega-regular properties on stochastic systems do not admit an equivalent additive reward structure.
A natural approach used in the literature (e.g., \cite{lazaric2008reinforcement}),
is to use heuristics for assigning additive rewards and then apply RL algorithms to obtain a policy. Unfortunately, there is no unique procedure for constructing these rewards and the learning does not necessarily converge to the optimal policy.
%
%
Due to all of these limitations, there is a need to provide data-driven algorithms that do not require any heuristics and have suitable convergence guarantees to policies that are optimal for satisfaction of temporal properties. 

\noindent\textbf{Related Works.}
In the last few years, researchers have started developing RL-based policy synthesis techniques in order to satisfy temporal properties.
There is a large body of literature in safe reinforcement learning (see e.g. \cite{garcia2015comprehensive,recht2018tour}). The problem of learning a policy to maximise the satisfaction probability of a temporal property was first introduced in 2014 \cite{brazdil2014verification,fu2014probably,sadigh2014learning}. 
The work \cite{brazdil2014verification} provides a heuristic-driven partial exploration of the model to find bounds for reachability probability.
The work \cite{fu2014probably} uses model-based RL in order to maximise the satisfaction probability of the property expressed as deterministic Rabin automaton (DRA). Given a Markov decision process (MDP) with unknown transition probabilities as the model of the system, the algorithms build a probably approximately correct MDP, which is then composed with the DRA for policy synthesis. The work \cite{sadigh2014learning} is limited to policies that generate traces satisfying the specification with probability one. The provided algorithm needs to compute all the transitions probabilities which in result requires a large memory usage.
This issue is partially addressed in \cite{wang2015temporal} by introducing an actor-critic algorithm that obtains transition probabilities only when needed in an approximate dynamic programming framework.

Satisfaction of LTL formulas can be checked on a class of automata called \emph{Limit-Deterministic B\"uchi Automata} (LDBA) \cite{courcoubetis1995complexity,Hahn2015LazyPM,sickert2016limit}. An implementation of a wide range of algorithms for translating LTL specifications to various types of automata is also available \cite{owl2018}.
%
The equivalent LDBA of \cite{sickert2016limit} is used in \cite{hasanbeig2018logically,hasanbeig2019reinforcement} to constrain the learning algorithm and is applied to an unknown finite MDP.
The work \cite{hahn2019omega} provides an RL-based policy synthesis for finite MDPs with unknown transition probabilities. It transforms the specification to an LDBA using \cite{Hahn2015LazyPM}, and then constructs a parameterised augmented MDP. It shows that the optimal policy obtained by RL for the reachability probability on the augmented MDP gives a policy for the MDP with a suitable convergence guarantee. 
In \cite{bozkurt2019control}, the authors utilise the LDBA representation, provide a path-dependent discounting mechanism for the RL algorithm, and prove convergence of their approach on finite MDPs when the discounting factor goes to one.

The literature on learning algorithms for formal synthesis on \emph{continuous-state} models is very limited. To the best of our knowledge, the only works developed for continuous-state stochastic models are \cite{hasanbeig2019logically,hosein2020,lavaei2020formal}.
The work \cite{lavaei2020formal} provides formal error bounds by discretising the space of the model, thus is only applicable to \emph{finite-horizon} properties.
The works \cite{hasanbeig2019logically,hosein2020} use respectively neural fitted Q-iteration and deep deterministic policy gradient(DDPG) approach without providing a proper formal convergence guarantee.
%
%
Our approach extends \cite{lavaei2020formal} to all LTL properties instead of finite-horizon properties and does not require any discretisation or knowledge of the continuity properties of the system.
Our approach is closely related to \cite{hahn2019omega} that discusses only \emph{finite-state} MDPs. We utilise the same technique and provide an example that shows the convergence guarantees of \cite{hahn2019omega} do not hold for all continuous-state MDPs but require an additional assumption on the model.
Our proofs are for general state spaces and do not rely on the properties of the bottom strongly-connected components of the MDP, thus simplify the ones in~\cite{hahn2019omega}. 
These proofs are
presented in the appendix.

\smallskip

\noindent\textbf{Main Contributions.}
We apply RL algorithms to \emph{continuous-state} stochastic systems using only paths of the system to find optimal policies satisfying an LTL specification.
We show that if a suitable assumption on the system holds, the formulated optimal average reward converges linearly to the true optimal satisfaction probability. We use negation of the specification and learn a lower bound on this satisfaction probability.  
To improve the performance of the learning on the constructed sparse reward function, we show how to construct a sequence of labelling functions based on the positive normal form of the LTL specification and use them for guiding the RL algorithm in learning the policy and its associated value function. This sequential learning is able to find policies for our case studies in less than $1.5$ hours but direct learning does not converge in 24 hours.

\smallskip

\noindent\textbf{Organisation.}
Section~\ref{sec:Prel_Prob} recalls definition of controlled Markov processes (CMPs) as the unknown model.
We also give linear temporal logic, limit-deterministic automata, and the problem statement in the same section. Section~\ref{sec:augmented_reach} gives construction of the augmented CMP and the product CMP. It establishes the relation between the reachability on the augmented CMP and the LTL satisfaction on the original CMP.
Section~\ref{sec:learning} gives the reward function for reachability on the augmented CMP that can be used by RL algorithms. It also gives a procedure for guiding the learning task via a sequence of labelling functions.
Finally, Section~\ref{sec:Case_studies} illustrates our approach on two case studies, a 4-dim cart-pole system and 6-dim boat driving problem.

\section{Preliminaries and Problem Statement}
\label{sec:Prel_Prob}


We consider a probability space $(\Omega,\mathcal F_{\Omega},P_{\Omega})$,
where $\Omega$ is the sample space,
$\mathcal F_{\Omega}$ is a sigma-algebra on $\Omega$ comprising subsets of $\Omega$ as events,
and $P_{\Omega}$ is a probability measure that assigns probabilities to events.
We assume that random variables introduced in this article are measurable functions of the form $X:(\Omega,\mathcal F_{\Omega})\rightarrow (S_X,\mathcal F_X)$ from the measurable space $(\Omega,\mathcal F_{\Omega})$ to a measurable space $(S_X,\mathcal F_X)$.
Any random variable $X$ induces a probability measure on its space $(S_X,\mathcal F_X)$ as $Prob\{A\} = P_{\Omega}\{X^{-1}(A)\}$ for any $A\in \mathcal F_X$.
We often directly discuss the probability measure on $(S_X,\mathcal F_X)$ without explicitly mentioning the underlying sample space and the function $X$ itself.

A topological space $S$ is called a Borel space if it is homeomorphic to a Borel subset of a Polish space (i.e., a separable and completely metrisable space).
Examples of a Borel space are the Euclidean spaces $\mathbb R^n$, its Borel subsets endowed with a subspace topology, as well as hybrid spaces of the form $Q\times \mathbb R^n$ with $Q$ being a finite set.
Any Borel space $S$ is assumed to be endowed with a Borel sigma-algebra, which is
denoted by $\mathcal B(S)$. We say that a map $f : S\rightarrow Y$ is measurable whenever it is Borel measurable.
We denote the set of non-negative integers by $\mathbb N := \{0,1,2,\ldots\}$ and the empty set by $\emptyset$.

\subsection{Controlled Markov Processes}
\label{subsec:model}
Controlled Markov processes (CMPs) are a natural choice for physical systems that have three main features: an uncountable state space that can be continuous or hybrid, control inputs to be designed, and inputs in the form of disturbance which have certain probabilistic behaviour \cite{dynkin1979controlled}.

We consider CMPs in discrete time defined over a general state space, 
characterised by the tuple
$\CMP =\left(\StS, \IS,\{\IS(s)|s\in\StS\}, \Ker\right),$
where $\StS$ is a Borel space as the state space of the CMP.
We denote by $(\StS, \mathcal B (\StS))$ the measurable space
with $\mathcal B (\StS)$ being the Borel sigma-algebra on the state space.
$\IS$ is a Borel space as the input space of the CMP.
The set $\{\IS(s)|s\in\StS\}$ is a family of non-empty measurable subsets of $\IS$ with the property that
$\mathcal K :=\{(s,u): s\in\StS, u\in\IS(s)\}$
is measurable in $\StS\times\IS$. Intuitively, $\IS(s)$ is the set of inputs that are feasible at state $s\in\StS$.
$\Ker:\mathcal B(\StS)\times \StS\times \IS\rightarrow[0,1]$,
is a conditional stochastic kernel that assigns to any $s \in \StS$ and $u\in\IS(s)$ a probability measure $\Ker(\cdot | s,u)$
on the measurable space
$(\StS,\mathcal B(\StS))$
so that for any set $A \in \mathcal B(\StS), P_{s,u}(A) = \int_A T_s (ds|s,u)$, 
where $P_{s,u}$ denotes the conditional probability $P(\cdot|s,u)$.
\tikzset{
    -{Latex[length=0.01mm,width=0.01mm]},auto,node distance =40 mm and 40 mm,semithick,
}
\definecolor{cwqctq}{rgb}{0.78,0.047,0.188}
\definecolor{qqtfxs}{rgb}{0,0.25,0.45}
\begin{figure}
\centering
\begin{tikzpicture}[line cap=round,line join=round,>=triangle 45,x=1cm,y=1cm, scale = .5]
\clip(-11,-1.3) rectangle (11,8.5);
\draw [line width=2pt,color=qqtfxs,fill=qqtfxs,fill opacity=1] (-1-3,0.5) circle (0.5cm);
\draw [line width=2pt,color=qqtfxs,fill=qqtfxs,fill opacity=1] (1-3,0.5) circle (0.5cm);
\fill[line width=2pt,color=qqtfxs,fill=qqtfxs,fill opacity=1] (-2-3,3) -- (2-3,3) -- (2-3,1) -- (-2-3,1) -- cycle;
\fill[line width=2pt,color=cwqctq,fill=cwqctq,fill opacity=0.24] (4,4) -- (8,4) -- (8,0) -- (4,0) -- cycle;

\draw [line width=2pt,color=cwqctq,fill=cwqctq,fill opacity=1] (0-3,2) circle (0.23565694610875978cm);
\draw [line width=5.2pt,color=cwqctq] (0-3,2)-- (0.6436459999999945-3,5.94564);
\draw[<->,>=stealth',semithick] (-1.8-3,7) arc[radius=5.31, start angle=110, end angle=70];
\draw [-, line width=0.5pt, dashed] (0-3,2) -- (1.8-3,7);
\draw [-, line width=0.5pt, dashed] (0-3,2) -- (-1.8-3,7);
\draw [<->, line width=0.5pt, dashed] (-10,-0.5) -- (10,-0.5);
\draw [-, line width=0.5pt] (-10,-0.8) -- (-10,-0.2);
\draw [-, line width=0.5pt] (10,-0.8) -- (10,-0.2);
\draw [line width=2pt] (-11,0)-- (11,0);
\draw [->, line width=0.6pt] (-3.5-3,2)-- (-2-3,2);
\draw (-2.9-3,2.3) node[anchor=center] {$u$};
\draw (6,1) node[anchor=center] {$A$};
\draw (0,-1) node[anchor=center] {$C_1$};
\draw (-3,8) node[anchor=center] {$C_2$};
\end{tikzpicture}
\caption{Cart-pole system with a 4-dim state space. It should stay within the limits specified by $C_1$, always keep the pole upright in the range $C_2$, and reach the region $A$.}
\label{cartpole}
\end{figure}
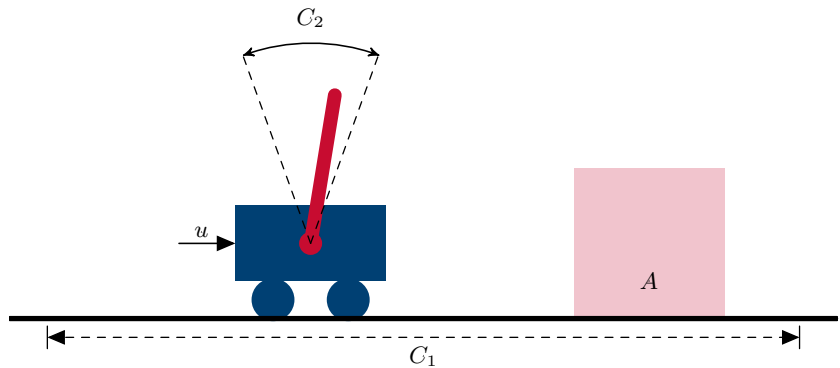

\begin{example}
\label{example1}
Consider the cart-pole in Figure~\ref{cartpole}. The cart moves along a line in either direction. The states are position $s^1$, velocity $s^2$, pole's angle $s^3$, and the angular velocity $s^4$. The input $u_n$ is the force applied to the cart at time step $n$. Its dynamics in discrete time are according to the following 4-dim difference equation:
\begin{equation}
\begin{cases}
s^1_{n+1} = s^1_n+\Delta s^2_n\\
s^2_{n+1} = s^2_n+\Delta a_3\\
s^3_{n+1} = s^3_n+\Delta s^2_n\\
s^4_{n+1} = s^4_n+\Delta a_2 + \eta_n,
    \end{cases}
    \text{ with }
    \begin{cases}
    a_3  := a_1 - \dfrac{l  a_2  \cos(s^3_n) }{(M+m)}\\
    a_2 := \dfrac{g  \sin(s^3_n) - \cos(s^3_n) a_1}{l(\frac{4}{3} - m  (\cos(s^3_n))^2 / (M+m))}\\
    a_1 := \dfrac{u_n + l  (s^4_n)^2 \sin(s^3_n)}{M+m}.
\end{cases}
\end{equation}
$\Delta$ is the sampling time, $M$ is the mass of the cart, $m$ is the mass of the pole, $l$ is the half length of the pole, and $\eta_n$ models the disturbance. The cart has discrete input and can be either pushed to the left or right with a fixed value, $\IS=\{-F_{max}, F_{max}\}$. This input $u_n$ appears in $a_1$ that affects both $a_2$ and $a_3$.
Assuming that the disturbances are all independent with normal distribution $\mathcal{N}(\,\cdot\,;\,0,\sigma^2)$, this system is a CMP with $\StS = \mathbb R^4$, $\IS(s) = \IS$ for all $s\in\StS$, and kernel
\begin{equation*}
\begin{array}{ll}
   \Ker(d\bar s\,|\,s,u)= \mathcal{N}(\,d\bar s^4\,;\,s^4_n+& \Delta a_2\,,\,\sigma^2) \boldsymbol{\delta}(d\bar s^1\,;\,s^1_n+\Delta s^2_n) \\
   &\times  \boldsymbol{\delta}(d\bar s^2\,;\,s^2_n+\Delta a_3) \boldsymbol{\delta}(d\bar s^3\,;\,s^3_n+\Delta s^2_n),
\end{array}
\end{equation*}
where $\boldsymbol{\delta}(\cdot\,;\,a)$ is the Dirac delta measure centred at $a$ and $\mathcal{N}(\cdot\,;\,m,\sigma^2)$ is the normal probability measure with mean $m$ and variance $\sigma^2$.
\end{example}

\subsection{Semantics of Controlled Markov Processes}
The semantics of a CMP is characterised by its \emph{paths} or
executions, which reflect both the history of previous states of
the system and of implemented control inputs.
Paths are used to measure the performance of the system.

\begin{definition}
A \emph{finite path} of $\CMP$ is a sequence
$w_n = (s_0,u_0,\ldots, s_{n-1},u_{n-1},s_n)$, $n\in \mathbb N$,
where $s_i\in\StS$ are state coordinates and $u_i\in\IS(s_i)$ are control input coordinates of the path.
The space of all paths of length $n$ is denoted by $\Paths_n:=\mathcal K^n\times\StS$. 
Further, we denote projections by $w_n[i]:=s_i$ and $w_n(i):=u_i$.
An \emph{infinite path} of the CMP $\CMP$ is the sequence
$w = (s_0,u_0,s_1,u_1,\ldots),$
where $s_i\in\StS$ and $u_i\in\IS(s_i)$ for all $i\in\mathbb N$.
As above, let us introduce $w[i]:=s_i$ and $w(i):=u_i$.
The space of all infinite paths is denoted by $\Paths_\infty:=\mathcal K^\infty$.
\end{definition}

Given an infinite path $w$ or a finite path $w_n$, we assume below that $s_i$ and $u_i$
are their state and control coordinates respectively, unless otherwise
stated. For any infinite path $w\in \Paths_\infty$,
its $n$-prefix (ending in a state) $w_n$ is a finite path of length $n$, which we also call \emph{$n$-history}.
We are now ready to introduce the notion of control policy. 

\begin{definition}
\label{def:policy}
A \emph{policy} is a sequence $\rho = (\rho_0,\rho_1,\rho_2,\ldots)$ of universally measurable stochastic kernels $\rho_n$ \cite{BS96},
each defined on the input space $\IS$ given $\Paths_n$
and such that for all $w_n\in \Paths_n$ with $n\in\mathbb N$, $\rho_n(\IS(s_n)|w_n)=1$.
The set of all policies is denoted by~$\Pi$.
\end{definition}

Given a policy $\rho\in\Pi$ and a finite path $w_n\in \Paths_n$, the distribution of the next control input $u_n$ is given by
$\rho_n(\cdot|w_n)$ and is supported on $\IS(s_n)$ (i.e., the chance of selecting an invalid input at $s_n$ is zero).
For a CMP $\CMP$, any policy $\rho\in \Pi$ together with an initial probability measure $\alpha:\mathcal{B}(\StS)\rightarrow[0,1]$ of the CMP induce a unique probability measure on the canonical sample space of paths \cite{hll1996} denoted by $P_\alpha^\rho$ with the expectation $\mathbb E_\alpha^\rho$.
When the initial probability measure is supported on a single point,
i.e., $\alpha({s}) = 1$, we write $P_s^{\rho}$ and $\mathbb E_s^{\rho}$ in place of $P_\alpha^\rho$ and $\mathbb E_\alpha^\rho$, respectively.
We denote the set of probability measures on $(\StS,\mathcal B(\StS))$ by $\mathfrak D$.
Implementation of a general policy requires an infinite memory. In this work, we restrict our attention to the class of policies that depend on the paths via a \emph{finite memory}.

\begin{definition}
\label{def:memory_policy}
A \emph{finite-memory policy} for $\CMP$ is a tuple $\rho_f := (\hatS,\hat s_0,\Kerp,\Kero)$, where $\hatS$ is the state space of the policy,
$\hat s_0\in \hatS$ is the initial state,
$\Kerp:\hatS\times\StS\times\mathcal B(\hatS)\rightarrow[0,1]$ is the stochastic kernel for updating the state of the policy, and $\Kero:\hatS\times\StS\times\mathcal B(\IS)\rightarrow[0,1]$ is the output kernel such that $\Kero(\IS(s)\,|\,\hat s,s)=1$ for all $\hat s\in\hatS$ and $s\in\StS$. We denote the set of such policies by $\Pi_f\subset\Pi$.
\end{definition}
Note that the state space $\hatS$ could in general be any continuous or hybrid space. The policy has access to the current state $s_n$ of $\CMP$ and updates its own state $\hat s_n$ according to $\hat s_{n+1}\sim \Kerp(\cdot\,|\,\hat s_n,s_n)$. As we will see later in Lemma~\ref{lem:policy}, a finite $\hatS$ is sufficient for optimal satisfaction of LTL specifications.

There is a special class of policies called \emph{positional} that do not need a memory state as defined next.
\begin{definition}
A policy $\rho$ is \emph{positional} if there is a stochastic kernel $\Cont:\StS\times\mathcal B(\IS)\rightarrow[0,1]$ such that at any time $n\in\mathbb N$, the input $u_n$ is taken from the probability measure $\Cont(\cdot|s_n)$.
Namely, the output kernel $\Kero(\cdot|\hat s,s)$ in Definition~\ref{def:memory_policy} is independent of $\hat s$.
We denote the class of positional policies by $\Pi_p\subset\Pi_f$ and a positional policy just by the kernel
$\Cont\in\Pi_p$.
\end{definition}

Designing optimal finite-memory policies to satisfy a specification on $\CMP$ can be reduced to finding an optimal positional policy for satisfying a specification on an extended model $\CMP'$. This is formally proved in Section~\ref{sec:augmented_reach}. Next we define the class of specifications used in this paper. 

\subsection{Linear Temporal Logic}
Linear temporal logic (LTL) provides a high-level language for describing the desired behaviour of a process. Formulas in this logic are constructed inductively by using a set of atomic propositions and combining them via Boolean operators. 
Consider a finite set of atomic propositions $\AP$ that defines the alphabet $\alphabeth := 2^{\AP}$. Thus, each letter of this alphabet evaluates a subset of the atomic propositions as true. Composed as an infinite string, these letters form infinite words defined as
$\word=\wordt{0},\wordt{1},\wordt{2},\ldots\in\alphabeth^{\mathbb{N}}$.
These words are connected to paths of CMP $\CMP$ via a measurable labelling function $\Lab:\StS\rightarrow \alphabeth$ that assigns letters $\alpha =\Lab(s)$ to state $s\in\StS$. That is, infinite paths $w = (s_0,u_0,s_1,u_1,\ldots)$ are mapped to the set of infinite words $\alphabeth^{\mathbb N}$, as
$\word=\Lab(w) := (\Lab(s_0),\Lab(s_1),\Lab(s_2),\ldots)$.

\begin{definition}
	\label{def:LTL}
	An LTL formula over a set of atomic propositions $\AP$ is constructed inductively as
	\begin{equation}
	\label{eq:PNF}
	\psi ::=  \textsf{true} \,|\, \textsf{false} \,|\, p \,|\, \neg p  \,|\,\psi_1 \wedge \psi_2 \,|\, \psi_1 \vee \psi_2 \,|\, \nex \psi \,|\, \psi_1\until \psi_2 \,|\, \psi_1 \release \psi_2,\quad p\in \AP,
	\end{equation}
	with $\psi_1,\psi_2,\psi$ being LTL formulas.
	\end{definition}

Let ${\word}_n=(\wordt{n},\wordt{n+1},\wordt{n+2},\ldots)$ be a postfix of $\word$. The satisfaction relation is denoted by $ \word \vDash\psi$
(or equivalently $\word_0\vDash\psi$) and is defined recursively as follows
\begin{itemize}
\item $\word_n\vDash\textsf{true}$ always hold and $\word_n\vDash\textsf{false}$ does not hold.
\item An atomic proposition, $ \word_n\vDash   p$  for $ p\in \AP$ holds if $p \in\wordt{n}$.
\item A negation, $\word_n\vDash\neg p$, holds if $ \word_n\nvDash p$.
\item A logical conjunction, $\word_n\vDash \psi_1\wedge\psi_2$, holds
if $ \word_n\vDash \psi_1$ and $ \word_n\vDash \psi_2$.
\item A logical disjunction, $\word_n\vDash \psi_1\vee\psi_2$, holds
if $ \word_n\vDash \psi_1$ or $ \word_n\vDash \psi_2$.
\item A temporal next operator, $\word_n\vDash\nex\psi$,  holds if $\word_{n+1}\vDash \psi$.
\item A temporal until operator, $\word_n\vDash \psi_1\until\psi_2$,  holds if there exists an $i \in \mathbb{N}$ such that $\word_{n+i} \vDash \psi_2$, and for all
$j \in\mathbb{N}$, $0\leq j<i$, we have $\word_{n+j}\vDash \psi_1$.
\item A temporal release operator is dual of the until operator and is defied as $\word_n\vDash \psi_1 \release \psi_2$ if $\word_n\nvDash\neg \psi_1\until \neg\psi_2$. 
\end{itemize}

In addition to the aforementioned operators, we can also use \emph{eventually} $\lozenge$, and \emph{always}~$\square$ operators as $\lozenge\psi:=\left(\textsf{true} \until \psi\right)$ and $\square\psi:=\textsf{false}\release\psi$.

\begin{remark}
The above definition is the canonical form of LTL and is called \emph{positive normal form} (PNF), in which negations only occur adjacent to atomic propositions. If this is not the case, it is possible to construct an equivalent formula \cite[Theorem 5.24]{BK08} in the canonical form in polynomial time as a function of the length of the formula.
We utilise the canonical form in Section~\ref{subsec:sequetial} to construct a sequence of learning procedures that guides the optimal policy learning problem.
\end{remark}

\noindent\textbf{Example 1 (Continued).}
The cart in Figure~\ref{cartpole} should stay within the limits specified by $C_1$, always keep the pole upright in the range $C_2$, and reach the region $A$. 
We can express this requirement as the LTL specification
\begin{equation}
\label{eq:spec_example}
     \psi = \lozenge a \wedge \square (c_1\wedge c_2)
\end{equation}
with $\AP = \{a, c_1,c_2\}$ and the labelling function $\Lab$ with $a\in \Lab(s)$ if the cart is inside $A$, $c_1\in \Lab(s)$ if the cart is inside $C_1$, and $c_2\in \Lab(s)$ if the pole angle is inside the specified range of $C_2$.

\subsection{Problem Statement}
We are interested in the probability that an LTL specification $\psi$ can be satisfied by paths of a CMP $\CMP$ under different policies.
Suppose a CMP $\CMP =\left(\StS, \IS,\{\IS(s)|s\in\StS\}, \Ker\right)$, an LTL specification $\psi$ over the alphabet $\Sigma$, and a labelling function $\Lab:\StS\rightarrow\Sigma$ are given.
An infinite path $w = (s_0,u_0,s_1,u_1,\ldots)$ of $\CMP$ satisfies $\psi$ if the infinite word
$\word=\Lab(w)\in \alphabeth^{\mathbb N}$ satisfies $\psi$. We denote such an event by $\CMP\models\psi$ and will study the probability of the event.
\begin{remark}
\label{rem:labelling_notation}
In general, one should use the notation $\CMP\models_\Lab\!\psi$ to emphasise the role of labelling function $\Lab$ in the satisfaction of $\psi$ by paths of $\CMP$. We eliminate the subscript $\Lab$ with the understanding that it is clear from the context. We add the labelling function in Section~\ref{subsec:sequetial} when discussing multiple labelling functions for evaluation of $\CMP\models\psi$. 
\end{remark}
Given a policy $\rho\in\Pi_f$ and initial state $s\in\StS$, we define the satisfaction probability as
$f(s,\rho) := P_s^{\rho}(\CMP\models \psi)$,
and the supremum satisfaction probability
\label{eq:optimal_prob}
$f^\ast(s) := \sup_{\rho\in\Pi_f} P_s^{\rho}(\CMP\models \psi)$.

\begin{problem}[Synthesis for LTL]
	\label{prob:LTL}
	Given CMP $\CMP$, LTL specification $\psi$, and labelling function $\Lab$, find an optimal policy $\rho^\ast\in\Pi_f$ along with $f^\ast(s)$ s.t.\ $P_s^{\rho^\ast}(\CMP\models \psi) = f^\ast(s)$.
\end{problem}

Measurability of the set $\{\CMP\models\psi\}$ in the canonical sample space of paths under the probability measure $P_s^{\rho}$ is proved in \cite{TMKA17}. The function $f^\ast(s)$ is studied in \cite{TMKA17} with an approximation procedure presented in \cite{majumdar2019symbolic}. These works are for fully known $\CMP$ and only for \emph{B\"uchi conditions} where the system should visit a set $B\subset\StS$ infinitely often. This condition is denoted by $\psi = \square\lozenge B$.

\begin{problem}[Synthesis for B\"uchi Conditions]
	\label{prob:Buchi}
	Given $\CMP$, a set of accepting states $B\in\mathcal B(\StS)$, find an optimal positional policy $\rho^*\in\Pi_p$ along with $f^\ast(s)$ s.t.\ $P_s^{\rho^\ast}(\CMP\models \square\lozenge B) = f^\ast(s)$.
\end{problem}

\begin{remark}
We have restricted our attention to finite-memory policies in Problem~\ref{prob:LTL}. This is due to the fact that proving existence of an optimal policy $\rho^*\in\Pi$ is an open problem. We note that existence of $\epsilon$-optimal policies is already proved \cite{MS_game_93,FPS_games_18}.
We prove in Section~\ref{sec:augmented_reach} that Problems~\ref{prob:LTL} and \ref{prob:Buchi} are closely related: in order to find a solution for Problem~\ref{prob:LTL}, we can find a solution for Problems~\ref{prob:Buchi} on another CMP with an extended state space.
\end{remark}

\subsection{Limit-Deterministic B\"uchi Automata}
\label{subsec:LDBA}

Satisfaction of LTL formulas can be checked on a class of automata called \emph{Limit-Deterministic B\"uchi Automata} (LDBA) \cite{courcoubetis1995complexity,Hahn2015LazyPM,sickert2016limit,hahn2019omega}. Similar to \cite{hahn2019omega}, we use the translation of the specification to an LDBA that has one set of accepting transitions and is presented next. This translation is provided by \cite{Hahn2015LazyPM}. An implementation of a wide range of algorithms for translating LTL to various types of automata is also available \cite{owl2018}.



\begin{definition}[LDBA]
an LDBA is a tuple $\LDBA = (Q,\Sigma,\delta,q_{0}, \Acc)$, where 
$Q$ is a finite set of states,
$\Sigma$ is a finite alphabet,
$\delta: Q\times (\Sigma\cup\{\epsilon\})\to 2^{Q}$ is a partial transition function, 
$q_0\in Q$ is an initial state, and
$\Acc\subset Q\times\Sigma\times Q$ is a set of accepting transitions. The transition function $\delta$ is such that it is total for all $(q,\wordt{})\in  Q\times\Sigma$, i.e., $|\delta(q,\wordt{})|\le 1$ for all $\wordt{}\neq \epsilon$ and $q\in Q$.
Moreover, there is a partition $\{Q_N,Q_D\}$ for $Q$ such that
\begin{itemize}
\item $\delta(q,\epsilon)=\emptyset$ for all $q\in Q_D$, i.e., the $\epsilon$-transitions can only occur in $Q_N$. 
\item $\delta(q,\wordt{})\subset Q_D$ for all $q\in Q_D$ and $\wordt{}\in\Sigma$, i.e., the transitions starting in $Q_D$ remain in $Q_D$.
\item $\Acc\subset Q_D\times\Sigma\times Q_D$, the accepting transitions start only in $Q_D$.
\end{itemize}
\end{definition}

We can associate to an infinite word $\word = (\wordt{0},\wordt{1},\wordt{2},\ldots)\in (\Sigma\cup\{\epsilon\})^{\mathbb N}$, a path $r = (q_0,\wordt{0},q_1,\wordt{1},q_2,\ldots)$ to $\LDBA$ such that $q_0$ is the initial state of $\LDBA$ and $q_{n+1}\in \delta(q_n,\wordt{n})$ for all $n\in\mathbb N$. Such a path always exists when $\word\in \Sigma^{\mathbb N}$.
Let us denote by $inf(r)$ as the set of transitions $(q,\wordt{},q')$ appearing in $r$ infinitely often. We say the word $\word$ is accepted by $\LDBA$ if it has a path $r$ with $inf(r)\cap\Acc\neq\emptyset$. The \emph{accepting language} of $\LDBA$ is the set of words accepted by $\LDBA$ and is denoted by $\mathcal L(\LDBA)$.

\section{Augmented CMP with Reachability Specification}
\label{sec:augmented_reach}

In this section we discuss approximating solutions of Problems~\ref{prob:LTL} and \ref{prob:Buchi} using reachability specifications. This section contains one of the main contributions of the paper that is formulating Assumption~\ref{ass:bounded} and proving Theorems~\ref{thm:augmented1}-\ref{thm:augmented_lower_bound} and Lemma~\ref{lem:policy} for continuous-state CMPs.

\subsection{The Augmented CMP}\label{subsec:augmented_cmp}

Given $\CMP = \left(\StS, \IS,\{\IS(s)|s\in\StS\}, \Ker\right)$ and a set of accepting states $B\subset\StS$, 
we construct an augmented CMP $\CMP_\zeta = \left(\StS_\zeta, \IS,\{\IS_\zeta(s)|s\in\StS_\zeta\},\Ker^\zeta\right)$ that has an additional dummy state $\phi$,
$\StS_\zeta := \StS\cup\{\phi\}$ and
the same input space $\IS$.
The set of valid inputs $\IS_\zeta(s)$ is the same as $\IS(s)$ for all $s\in\StS$ and $\IS_\zeta(\phi) = \IS$. The stochastic kernel of $\CMP_\zeta$ is a modified version of $\Ker$ as $\Ker^\zeta(A|s,u) = [1-(1-\zeta)\mathbf{1}_B(s)]\Ker(A|s,u)$, $\Ker^\zeta(\phi|s,u) = (1-\zeta)\mathbf{1}_B(s)$, and $\Ker(\phi|\phi,u)=1$, for all $A\in\mathcal B(\StS)$, $s\in\StS$ and $u\in \IS_\zeta(s)$.
In words, $\Ker^\zeta$ takes the same $\Ker$, adds a sink state $\phi$, and for any accepting state $s\in B$, the process will jump to $\phi$ with probability $(1-\zeta)$. It also normalises the outgoing transition probabilities of accepting ones with $\zeta$.
We establish a relation between $\CMP$ and $\CMP_\zeta$ regarding satisfaction of B\"uchi conditions under the following assumption.
\begin{assumption}
\label{ass:bounded}
For $\CMP$ and a set $B$, define the random variable $\tau_B$ as the number of times the set $B$ is visited in paths of $\CMP$ conditioned on having it as a finite number. The quantity
$\tau^\ast_B:=\sup_{\rho} \mathbb E_s^\rho(\tau_B)$ is bounded for any $s\in\StS$.
\end{assumption}

\begin{theorem}
\label{thm:augmented1}
Given $\CMP$ satisfying Assumption~\ref{ass:bounded} and for any positional policy $\rho$ on $\CMP$, there is a positional policy $\bar\rho$ on $\CMP_\zeta$ such that
\begin{equation}
\label{eq:reach1}
P_s^{\bar\rho}(\CMP_\zeta\models\lozenge\phi) - (1-\zeta)\mathbb E_s^\rho(\tau_B) \le
    P_s^{\rho}(\CMP\models \square\lozenge B) \le  P_s^{\bar\rho}(\CMP_\zeta\models\lozenge\phi).
\end{equation}
For any $\bar\rho$ on $\CMP_\zeta$, there is $\rho$ on $\CMP$ such that the same inequality holds.
\end{theorem}

The above theorem shows that the probability of satisfying a B\"uchi condition with accepting set $B\subset\StS$ by $\CMP$ is upper bounded by the probability of reaching $\phi$ in $\CMP_\zeta$. It also establishes a lower bound but requires knowing $\mathbb E_s^\rho(\tau_B)$. 


Inequalities of Theorem~\ref{thm:augmented1} can be extended to optimal satisfaction probabilities as stated in the next theorem.
\begin{theorem}
\label{thm:augmented_optimal}
For any $\CMP$ satisfying Assumption~\ref{ass:bounded}, we have
\begin{equation}
\label{eq:reach_optimal}
    \sup_{\bar\rho}P_s^{\bar\rho}(\CMP_\zeta\models\lozenge\phi) - (1-\zeta)\tau^\ast_B \le \sup_{\rho} P_s^{\rho}(\CMP\models \square\lozenge B) \le \sup_{\bar\rho}P_s^{\bar\rho}(\CMP_\zeta\models\lozenge\phi).
\end{equation}
\end{theorem}

\begin{corollary}
\label{cor:limit}
Under Assumption~\ref{ass:bounded}, the optimal value $\sup_{\bar\rho} P_s^{\bar\rho}(\CMP_\zeta\models\lozenge\phi)$ converges to $\sup_{\rho} P_s^{\rho}(\CMP\models \square\lozenge B)$ from above when $\zeta$ converges to one from below, and the rate of convergence is at least linear with $(1-\zeta)$.
\end{corollary}

Next example highlights the need for Assumption~\ref{ass:bounded} on $\CMP$ to get the linear convergence. Such an assumption holds for all $\CMP$ with finite state spaces as used in \cite{hahn2019omega,bozkurt2019control} but it may not hold for $\CMP$ with infinite state spaces.

\begin{example}
Consider the $\CMP$ presented in Figure~\ref{fig:countable_MDP}, which has a countable state space $\{1,2,3,\ldots\}$ and the input space is singleton. $\CMP$ starts at state $s=2$. The state $1$ is absorbing. From any other state $n$, it jumps to state $1$ with probability $\frac{1}{n}$ and to state $(n+1)$ with probability $\frac{n-1}{n}$. Take the set of accepting states $B = \{3,4,5,\ldots\}$.
$\mathbb E_s^\rho(\tau_B)$ is unbounded for $\CMP$:
\begin{eqnarray*}
    \mathbb E_s^\rho(\tau_B) = \sum_{n=1}^\infty
    n\times \frac{1}{2}\times \frac{2}{3} \times \frac{3}{4}\times\cdots \frac{n}{n+1}\times \frac{1}{n+2} = \sum_{n=1}^\infty \frac{n}{(n+1)(n+2)}  = \infty.
\end{eqnarray*}
It can be easily verified that
\begin{align*}
& P_s^{\rho}(\CMP\models \square\lozenge B) = \frac{1}{2}\times \frac{2}{3}\times \frac{3}{4}\times\frac{4}{5}\times\cdots = 0\\
    & P_s^{\bar\rho}(\CMP_\zeta\models\lozenge\phi) = (1-\zeta)\left[1+\frac{1}{2}\zeta + \frac{1}{3}\zeta^2+ \frac{1}{4}\zeta^3+\ldots \right] = \frac{-(1-\zeta)\ln(1-\zeta)}{\zeta}.
\end{align*}
The left-hand side of inequality~\eqref{eq:reach_optimal} is still technically true for this $\CMP$ despite $\mathbb E_s^\rho(\tau_B) = \infty$, but the provided lower bound is trivial and does not give linear convergence mentioned in Corollary~\ref{cor:limit}.

\tikzset{
    -{Latex[length=0.01mm,width=0.01mm]},auto,node distance =40 mm and 40 mm,semithick,
    ->-/.style={decoration={
  markings,
  mark=at position .5 with {\arrow{>}}},postaction={decorate}}
}
\begin{figure}
    \centering
\begin{tikzpicture}[x=0.75pt,y=0.75pt,yscale=-.5,xscale=.5]

\draw   (15,81) .. controls (15,67.19) and (26.19,56) .. (40,56) .. controls (53.81,56) and (65,67.19) .. (65,81) .. controls (65,94.81) and (53.81,106) .. (40,106) .. controls (26.19,106) and (15,94.81) .. (15,81) -- cycle ;
\draw   (122,81) .. controls (122,67.19) and (133.19,56) .. (147,56) .. controls (160.81,56) and (172,67.19) .. (172,81) .. controls (172,94.81) and (160.81,106) .. (147,106) .. controls (133.19,106) and (122,94.81) .. (122,81) -- cycle ;
\draw   (285,246) .. controls (285,232.19) and (296.19,221) .. (310,221) .. controls (323.81,221) and (335,232.19) .. (335,246) .. controls (335,259.81) and (323.81,271) .. (310,271) .. controls (296.19,271) and (285,259.81) .. (285,246) -- cycle ;
\draw   (229,81) .. controls (229,67.19) and (240.19,56) .. (254,56) .. controls (267.81,56) and (279,67.19) .. (279,81) .. controls (279,94.81) and (267.81,106) .. (254,106) .. controls (240.19,106) and (229,94.81) .. (229,81) -- cycle ;
\draw   (336,81) .. controls (336,67.19) and (347.19,56) .. (361,56) .. controls (374.81,56) and (386,67.19) .. (386,81) .. controls (386,94.81) and (374.81,106) .. (361,106) .. controls (347.19,106) and (336,94.81) .. (336,81) -- cycle ;
\draw   (443,81) .. controls (443,67.19) and (454.19,56) .. (468,56) .. controls (481.81,56) and (493,67.19) .. (493,81) .. controls (493,94.81) and (481.81,106) .. (468,106) .. controls (454.19,106) and (443,94.81) .. (443,81) -- cycle ;
\draw   (550,81) .. controls (550,67.19) and (561.19,56) .. (575,56) .. controls (588.81,56) and (600,67.19) .. (600,81) .. controls (600,94.81) and (588.81,106) .. (575,106) .. controls (561.19,106) and (550,94.81) .. (550,81) -- cycle ;
\draw   (40,56) .. controls (76.94,35.31) and (112.42,35.97) .. (145.49,55.11) ;
\draw [shift={(147,56)}, rotate = 210.84] [color={rgb, 255:red, 0; green, 0; blue, 0 }  ][line width=0.75]    (10.93,-3.29) .. controls (6.95,-1.4) and (3.31,-0.3) .. (0,0) .. controls (3.31,0.3) and (6.95,1.4) .. (10.93,3.29)   ;

\draw    (147,56) .. controls (183.94,35.31) and (219.42,35.97) .. (252.49,55.11) ;
\draw [shift={(254,56)}, rotate = 210.84] [color={rgb, 255:red, 0; green, 0; blue, 0 }  ][line width=0.75]    (10.93,-3.29) .. controls (6.95,-1.4) and (3.31,-0.3) .. (0,0) .. controls (3.31,0.3) and (6.95,1.4) .. (10.93,3.29)   ;

\draw    (254,56) .. controls (290.94,35.31) and (326.42,35.97) .. (359.49,55.11) ;
\draw [shift={(361,56)}, rotate = 210.84] [color={rgb, 255:red, 0; green, 0; blue, 0 }  ][line width=0.75]    (10.93,-3.29) .. controls (6.95,-1.4) and (3.31,-0.3) .. (0,0) .. controls (3.31,0.3) and (6.95,1.4) .. (10.93,3.29)   ;

\draw    (361,56) .. controls (397.94,35.31) and (433.42,35.97) .. (466.49,55.11) ;
\draw [shift={(468,56)}, rotate = 210.84] [color={rgb, 255:red, 0; green, 0; blue, 0 }  ][line width=0.75]    (10.93,-3.29) .. controls (6.95,-1.4) and (3.31,-0.3) .. (0,0) .. controls (3.31,0.3) and (6.95,1.4) .. (10.93,3.29)   ;

\draw    (468,56) .. controls (504.94,35.31) and (540.42,35.97) .. (573.49,55.11) ;
\draw [shift={(575,56)}, rotate = 210.84] [color={rgb, 255:red, 0; green, 0; blue, 0 }  ][line width=0.75]    (10.93,-3.29) .. controls (6.95,-1.4) and (3.31,-0.3) .. (0,0) .. controls (3.31,0.3) and (6.95,1.4) .. (10.93,3.29)   ;

\draw    (40,106) -- (283.26,245.01) ;
\draw [shift={(285,246)}, rotate = 209.74] [color={rgb, 255:red, 0; green, 0; blue, 0 }  ][line width=0.75]    (10.93,-3.29) .. controls (6.95,-1.4) and (3.31,-0.3) .. (0,0) .. controls (3.31,0.3) and (6.95,1.4) .. (10.93,3.29)   ;

\draw [->]   (147,106) -- (308.37,219.85) ;
\draw [shift={(310,221)}, rotate = 215.2] [color={rgb, 255:red, 0; green, 0; blue, 0 }  ][line width=0.75]    (10.93,-3.29) .. controls (6.95,-1.4) and (3.31,-0.3) .. (0,0) .. controls (3.31,0.3) and (6.95,1.4) .. (10.93,3.29)   ;

\draw    (254,106) -- (309.12,219.2) ;
\draw [shift={(310,221)}, rotate = 244.04000000000002] [color={rgb, 255:red, 0; green, 0; blue, 0 }  ][line width=0.75]    (10.93,-3.29) .. controls (6.95,-1.4) and (3.31,-0.3) .. (0,0) .. controls (3.31,0.3) and (6.95,1.4) .. (10.93,3.29)   ;

\draw    (361,106) -- (310.81,219.17) ;
\draw [shift={(310,221)}, rotate = 293.92] [color={rgb, 255:red, 0; green, 0; blue, 0 }  ][line width=0.75]    (10.93,-3.29) .. controls (6.95,-1.4) and (3.31,-0.3) .. (0,0) .. controls (3.31,0.3) and (6.95,1.4) .. (10.93,3.29)   ;

\draw    (468,106) -- (311.62,219.82) ;
\draw [shift={(310,221)}, rotate = 323.95] [color={rgb, 255:red, 0; green, 0; blue, 0 }  ][line width=0.75]    (10.93,-3.29) .. controls (6.95,-1.4) and (3.31,-0.3) .. (0,0) .. controls (3.31,0.3) and (6.95,1.4) .. (10.93,3.29)   ;

\draw    (575,106) -- (336.73,244.99) ;
\draw [shift={(335,246)}, rotate = 329.74] [color={rgb, 255:red, 0; green, 0; blue, 0 }  ][line width=0.75]    (10.93,-3.29) .. controls (6.95,-1.4) and (3.31,-0.3) .. (0,0) .. controls (3.31,0.3) and (6.95,1.4) .. (10.93,3.29)   ;

\draw    (575,56) .. controls (611.94,35.31) and (647.42,35.97) .. (680.49,55.11) ;
\draw [shift={(682,56)}, rotate = 210.84] [color={rgb, 255:red, 0; green, 0; blue, 0 }  ][line width=0.75]    (10.93,-3.29) .. controls (6.95,-1.4) and (3.31,-0.3) .. (0,0) .. controls (3.31,0.3) and (6.95,1.4) .. (10.93,3.29)   ;

\draw    (311.06,272.5) .. controls (359.44,335.85) and (257.53,335.01) .. (309.21,271.96) ;
\draw [shift={(310,271)}, rotate = 489.89] [color={rgb, 255:red, 0; green, 0; blue, 0 }  ][line width=0.75]    (10.93,-3.29) .. controls (6.95,-1.4) and (3.31,-0.3) .. (0,0) .. controls (3.31,0.3) and (6.95,1.4) .. (10.93,3.29)   ;

\draw (40,81) node    {$2$};
\draw (468,81) node    {$6$};
\draw (361,81) node    {$5$};
\draw (254,81) node    {$4$};
\draw (147,81) node    {$3$};
\draw (575,81) node    {$7$};
\draw (310,246) node    {$1$};
\draw (629,81) node    {$\dotsc $};
\draw (89,22) node    {$\frac{1}{2}$};
\draw (442,145) node    {$\frac{1}{6}$};
\draw (360,145) node    {$\frac{1}{5}$};
\draw (256,145) node    {$\frac{1}{4}$};
\draw (170,145) node    {$\frac{1}{3}$};
\draw (71,145) node    {$\frac{1}{2}$};
\draw (628,22) node    {$\frac{6}{7}$};
\draw (520,22) node    {$\frac{5}{6}$};
\draw (411,22) node    {$\frac{4}{5}$};
\draw (307,22) node    {$\frac{3}{4}$};
\draw (197,22) node    {$\frac{2}{3}$};
\draw (546,145) node    {$\frac{1}{7}$};
\draw (310,337) node    {$1$};
\end{tikzpicture}
    \caption{A CMP with space $\{1,2,3,\ldots\}$, a single input and accepting states $B = \{3,4,5,\ldots\}$. Its augmented CMP $\CMP_\zeta$ does not show convergence with a linear rate.}
    \label{fig:countable_MDP}
\end{figure}
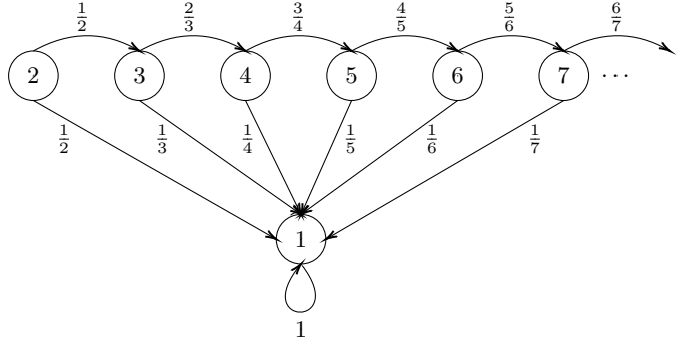
\end{example}

\begin{remark}
The lower bound in \eqref{eq:reach_optimal} is useful for showing linear convergence when $\zeta\rightarrow 1^-$, but it is not beneficial for learning purposes since the computation of $\tau^\ast_B$ requires knowing the structure of the underlying unknown transition kernel $\Ker$. In the next subsection, we utilise Theorem~\ref{thm:augmented_optimal} to give a lower bound independent of $\tau^\ast_B$. We also demonstrate convergence experimentally in the case study section. 
\end{remark}

\subsection{The Product CMP}
\label{subsec:product_cmp}
The product of a CMP and an LDBA is used in the literature, e.g., \cite{hahn2019omega,bozkurt2019control,hasanbeig2019reinforcement} for finite state spaces. We provide this construction for continuous-state CMPs.

\begin{definition}
The product CMP $\CMP^\otimes = (\StS^\otimes, \IS^\otimes, \{\IS^\otimes(x)|x \in \StS^\otimes\}, \mathcal{T}^\otimes_x)$ of an CMP $(\StS, \IS, \{\IS(s)|s \in \StS\}, \mathcal{T}_s)$ and an LDBA $\mathcal A= (Q,\Sigma,\delta, q_{0}, \Acc)$ is defined as follows:
$\StS^\otimes:=S\times Q$ is the set of states, $\IS^\otimes := \IS\cup A^\epsilon$ with $A^\epsilon: =\{\epsilon_q|q\in Q\}$ is the set of actions. The valid input sets are $\IS^\otimes(s,q)= \IS(s)$ if $\delta(q,\epsilon)=\emptyset$ and $\IS^\otimes(s,q)= \epsilon_{q'}$ if $q'\in \delta(q,\epsilon)$.
The stochastic kernel is defined as
\begin{equation*}
    \mathcal{T}^\otimes_x(A\times\{q'\}|s,q,u) :=
    \begin{cases}
    \Ker(A|s,u) & \text{if } q' = \delta(q,\Lab(s) \text{ and } u\in\IS(s)\\
    \mathbf 1_A(s) & \text{if } q' = \delta(q,\epsilon) \text{ and } u = \epsilon_{q'}\\
    0, & \text{ otherwise,}
    \end{cases}
\end{equation*}
where $\mathbf 1_A(s)$ is the indicator function of the set $A$.
\end{definition}
Any distribution $\alpha:\mathcal{B}(\StS)\rightarrow[0,1]$ for the initial state of $\CMP$ induces an initial distribution $\alpha^\otimes:\mathcal{B}(\hatS)\rightarrow[0,1]$ with $\alpha^\otimes(A\times\{q\}) := \alpha(A)$ for any $A\in\mathcal B(\StS)$ and $q=q_0$, and zero otherwise.
The set of accepting states in the product CMP $\CMP^\otimes$ is
\begin{equation}
\label{eq:accepting}
    \Acc^\otimes = \left\{(s,q)\,|\, (q,\Lab(s),q')\in\Acc, q' = \delta(q,\Lab(s))\right\}.
\end{equation}
We say the path $w^\otimes$ of $\CMP^\otimes$ satisfies the B\"uchi condition $\psi_B$ if the number of states in $\Acc^\otimes$ visited by the path is not finite (the set is visited infinitely often).



\begin{lemma}
\label{lem:policy}
Any positional policy on $\CMP^\otimes$ can be translated into a finite-memory policy for $\CMP$ that has a finite state space equal to the space of the LDBA $\mathcal A$. Moreover,
the class of finite-memory policies are sufficient for solving Problem~\ref{prob:LTL} if an optimal policy exists.
\end{lemma}

Due to Lemma~\ref{lem:policy}, we focus in the next section on finding positional policies for the product CMP using reinforcement learning. Next theorem is one of the main contributions of the paper that formalises a lower bound on the optimal satisfaction probability.

\begin{theorem}
\label{thm:augmented_lower_bound}
For any $\CMP$, specification $\psi$, labelling function $\Lab$, and any $s\in\StS$,
\begin{equation}
\label{eq:lower_bound}
    1-\inf_{\bar\rho}P_{s,q_0}^{\bar\rho}(\CMP^\otimes_{1\zeta}\models\lozenge\phi) \le \sup_{\rho} P_s^{\rho}(\CMP\models \psi) \le \sup_{\bar\rho}P_{s,q_0}^{\bar\rho}(\CMP^\otimes_{2\zeta}\models\lozenge\phi),
\end{equation}
where $\CMP^\otimes_{1\zeta}$ and $\CMP^\otimes_{2\zeta}$ are the augmented CMPs constructed for the products of $\CMP$ with $\LDBA_{\neg\psi}$ and $\LDBA_{\psi}$, respectively.
\end{theorem}

In the next section, we focus on the computation of the right-hand side of~\eqref{eq:lower_bound} using RL. The left-hand side is computed similarly.

\section{Reinforcement Learning for Policy Synthesis}
\label{sec:learning}

This section contains another main contributions of the paper that is using relaxed versions of the LTL specification in learning a policy. We have shown that Problem~\ref{prob:LTL} can be reduced to Problem~\ref{prob:Buchi} on a product CMP, which then can be approximated using reachability objectives as shown in \eqref{eq:lower_bound}.
The reachability probability is an average reward criterion
\begin{equation}
\label{eq:average_reward}
     P_s^{\bar\rho}(\CMP_\zeta\models\lozenge\phi) = \lim_{N\rightarrow\infty}\frac{1}{N+1}\mathbb E_s^{\bar\rho}\sum_{n=0}^N R(s_n),
\end{equation}
with the reward function $R:\StS_\zeta\rightarrow\mathbb R$ defined as $R(s)=1$ for $s=\phi$ and $R(s)=0$ otherwise. It can alternatively be written with a total (undiscounted) additive reward criterion by assigning reward one to the first visit of the $\phi$ and zero otherwise. Both cases can be computed by RL algorithms whenever the model of the CMP is not known or is hard to analyse. Any off-the-shelf RL algorithm for continuous systems can be used to learn a policy.
Note that for a general LTL specification, the reward function $R$ is state dependent on the product CMP, but it becomes path dependent when interpreted over the original CMP through the LDBA of the specification.   

\paragraph{Advantage Actor-Critic RL.}
RL algorithms are either value based or policy based. In value-based RL, the algorithm tries to maximise a value function that is a mapping between a state-input pair and a value. 
Policy-based RL tries to find the optimal policy without using a value function.
The policy-based RL has better convergence and effectiveness on high dimensions or continuous state spaces, while value-based RL is more sample efficient and steady. The intersection between these two categories is the actor-critic RL, where the goal is to optimise the policy and the value function together. It optimises the policy and value function as a function of state.
We use in this paper the \emph{Advantage Actor-Critic RL} (A2C) \cite{DeepRL16} that takes the value function as a baseline. It makes the cumulative reward smaller by subtracting it with the baseline, thus have smaller gradients and more stable updates. It works better in comparison with other actor-critic RL in terms of the stability of the learning process and lower variance. An implementation of A2C is available in MATLAB. We have taken this implementation and adapted it to be applicable to the augmented CMP $\CMP^\otimes_\zeta$. A pseudo algorithm of our approach based on the A2C is provided in the
appendix (Section~\ref{sec:A2C}, Alg.~\ref{al1}).
The algorithm
computes the parameters $\theta^{\mu}$ of an Actor network $\mu(s,q|\theta^{\mu})$ for the policy and the parameters $\theta^{\mathcal Q}$ of a Critic network $\mathcal Q(s,q|\theta^{\mathcal Q})$ for the value generated by applying that policy.

\subsection{Specification-Guided Learning}
\label{subsec:sequetial}
The reward function $R$ used in \eqref{eq:average_reward} is sparse and it slows down the learning. To improve the learning performance, we give an algorithm that sequentially trains the Actor and Critic networks and guides the learning process by a sequence of labelling functions defining satisfaction of the specification with different relaxation degrees. 
This sequential training has a similar spirit as the approach of \cite{lazaric2008reinforcement}. The novelty of our algorithm is in constructing a sequence of labelling functions that automatically encode the satisfaction relaxation, thus requires Actor and Critic networks with fixed structures.   

\paragraph{Relaxed labelling functions.}We denote the elements of the alphabet by $\Sigma=\{\Sigma_1,\ldots,\Sigma_m\}$.
The labelling function $\Lab:\StS\rightarrow\Sigma$ induces a partition of the state space $\{S_1,S_2,\ldots,S_m\}$ such that $S_i:=\Lab^{-1}(\Sigma_i)$, $\StS = \cup_{i=1}^n S_i$, and $S_i\cap S_j = \emptyset$ for all $i\ne j$.
Define the \emph{$r$-expanded} version of a set $S\subset\StS$ by
\begin{equation}
    S^{+r}:=\{s\in \StS\,|\, \exists s'\in S \text{ with } \|s-s'\|_\infty\le r\},
\end{equation}
for any $r\ge 0$, where $\|\cdot\|_\infty$ is the infinity norm. Define the \emph{$r$-relaxed} labelling function $\Lab_r:\StS\rightarrow 2^\Sigma$ with
\begin{equation}
\label{eq:relaxed_label}
    \Lab_r(s) := \{\Sigma_i\,|\, \Lab(S_i) = \Sigma_i \text{ and } s\in S_i^{+r}\},\quad \text{ for all }s\in\StS.
\end{equation}

\begin{theorem}
\label{thm:sequence_r}
The relaxed labelling functions $\Lab_r$ are monotonic with respect to~$r$, i.e., for any $0\le r\le r'$ and $\Lab$, we have $\{\Lab(s)\} = \Lab_{0}(s) \subset\Lab_{r}(s)\subset \Lab_{r'}(s)$. 
\end{theorem}

\paragraph{Specification interpreted over $\Sigma$.}We interpret the specification $\psi$ over the letters in  $\Sigma$ instead of the atomic propositions in $\AP$. For this, we take the PNF form of $\psi$ and replace an atomic proposition $p$ by $\vee_{i}\{\Sigma_i\,|\,p\in\Sigma_i\}$. We also replace $\neg p$ by $\vee_{i}\{\Sigma_i\,|\,p\notin\Sigma_i\}$.
Let us denote this specification in PNF with the letters $\{\Sigma_1,\ldots,\Sigma_m\}$ treated as atomic propositions $\bar \psi$. We can construct its associated LDBA $\bar{\LDBA}_\psi$ as discussed in Section~\ref{subsec:LDBA},
\begin{equation}
    \label{eq:extended_LDBA}
    \bar{\LDBA}_\psi := (\bar Q,2^\Sigma,\bar\delta,\bar q_{0},\overline{\Acc}).
\end{equation}

\begin{theorem}
\label{thm:relaxed}
For any $0\le r\le r'$ and $\Lab$, we have
\begin{equation}
\label{eq:sequence_r}
\left\{\CMP\models_{\Lab}\!\psi\right\} = \left\{\CMP\models_{\Lab_0}\!\bar\psi\right\}\subset \left\{\CMP\models_{\Lab_r}\!\bar\psi\right\} \subset \left\{\CMP\models_{\Lab_{r'}}\bar\psi\right\},
\end{equation}
where $\Lab_{r}$ is the $r$-relaxed labelling function defined in~\eqref{eq:relaxed_label}, and $\bar\psi$ is the specification $\psi$ in PNF and interpreted over $\Sigma$. 
\end{theorem}
%

A pseudo algorithm for the specification-guided learning is provided in Alg.~\ref{al2} that is based on repeatedly applying an RL algorithm to $\CMP$ using a sequence of $r$-relaxed labelling functions.
The algorithm starts by applying Actor-Critic RL
(e.g., Alg.~\ref{al1})
to the most relaxed labelling function $\Lab_{r_{\mathsf m}}$. Then it repeatedly fixes the actor network (the policy) by setting its learning rate to zero (Step~\ref{step:learn0}), runs Actor-Critic RL on the next most relaxed labelling function to update the Critic network that gives the total reward (Step~\ref{step:onlycritic}), and uses these two networks as initialisation for running Actor-Critic RL to optimise both Actor and Critic networks (Step~\ref{step:train_both}).
    \begin{remark}
    The main feature of Alg.~\ref{al2} is that the structure of the LDBA $\bar{\LDBA}_\psi$ is fixed through the entire algorithm and only the labelling function (thus the reward function) is changed in each iteration.
    \end{remark}
We presented Alg.~\ref{al2} for the computation of the right-hand side of~\eqref{eq:lower_bound}. The lower bound in \eqref{eq:lower_bound} is computed similarly. The only difference is that the LDBA is constructed using $\neg\psi$. The reward function should assign zero to $\phi$ and one to all other states. The $r$-relaxed labelling functions in \eqref{eq:relaxed_label} can be used for guiding the computation of the lower bound.  
    
\begin{algorithm2e}[t]
\label{al2}
\DontPrintSemicolon
\SetKw{return}{return}
\SetKwRepeat{Do}{do}{while}
\SetKwInOut{Input}{input}
\SetKwInOut{Output}{output}
\SetKwFor{Loop}{Loop}{}{}
\caption{Specification-Guided Learning}
\begin{small}
    \Input{
    CMP $\CMP$ as a black box, specification $\psi$, labelling function $\Lab:\StS\rightarrow\Sigma$}
    \Output{Actor network $\mu(s,q|\theta^{\mu})$ and Critic network $\mathcal Q(s,q|\theta^{\mathcal Q})$}
    Select \textbf{hyper-parameters} $r_\mathsf{m}> r_{\mathsf{m}-1}>\ldots r_1 > r_0 = 0$\;
    Compute \emph{$r$-relaxed} labelling functions $\Lab_{r_i}:\StS\rightarrow 2^\Sigma$ according to \eqref{eq:relaxed_label}\;
    Compute LDBA $\bar{\LDBA}_\psi$ as discussed for \eqref{eq:extended_LDBA}\;
    Run the Actor-Critic RL 
    (Alg.~\ref{al1})
    with ($\CMP$, $\bar{\LDBA}_\psi$, $\Lab_{r_{\mathsf m}}$) to get Actor and Critic networks $\mu(s,q|\theta^{\mu})$ and $\mathcal Q(s,q|\theta^{\mathcal Q})$ \;
    \For{$i=\mathsf m$ \textbf{to} $1$}
    {   
    Fix parameters $\theta^{\mu}$ of the Actor network by setting its learning rate to zero\label{step:learn0}\;
    Run Actor-Critic RL 
    (Alg.~\ref{al1})
    with $\Lab_{r_{i-1}}$ to train only the Critic network\label{step_init}\label{step:onlycritic}\;
        Change the learning rate of Actor back to normal\;
        Run  Actor-Critic RL 
        (Alg.~\ref{al1})
        with $\Lab_{r_{i-1}}$ and initial parameters obtained in Steps~\ref{step:learn0} and \ref{step_init}\label{step:train_both}\;
    }
\end{small}
\end{algorithm2e}

\section{Case Studies}
\label{sec:Case_studies}

To demonstrate our model-free policy synthesis method, we first apply it to the cart-pole system of Example~\ref{example1} and then discuss the results on a $6$-dim boat driving problem.
Note that it is not possible to compare our approach with \cite{lavaei2020formal} that only handles finite-horizon specifications. Also, the approach of \cite{hasanbeig2019logically,hosein2020} maximises the frequency of visiting a sequence of sets of accepting transitions and does not come with formal convergence or lower-bound guarantees.

Our algorithms are implemented in MATLAB R2019a on a 64-bit machine with an Intel Core(TM) i7 CPU
at 3.2 GHz and 16 GB RAM.


\subsection{Cart-Pole System}
We use negation of the specification \eqref{eq:spec_example} to learn a lower bound on the optimal satisfaction probability. We set the safe interval $C_2 = [-12^\circ,12^\circ]$ for the angle, safe range $C_1 = [-1,1]$ and reach set $A = [0.4,1]$ in meters for the location. We first directly apply A2C RL to the specification \eqref{eq:spec_example} and set the timeout of $24$ hours. The RL does not converge to a policy within this time frame. Note that it is a very challenging task to keep the pole upright and at the same time move the cart to reach the desired location.

We then apply Alg.~\ref{al2} by using the expanded sets
$A^{+i} = [\alpha_i,1]$ with $\alpha_i\in\{-1,0.01,0.4\}$
for defining the relaxed labelling functions $\Lab_i$.
We select the Actor network to have $7$ inputs ($4$ real states and $3$ discrete states of the automaton) and $2$ outputs. It also has two fully-connected hidden layers each with $7$ nodes.
The Critic network has the same number of of inputs as Actor network, one output, and one fully-connected layer with $7$ nodes. We also set $\zeta = 0.999$, learning rate $8\times 10^{-4}$, and episode horizon $N = 500$.

\begin{figure}
  \centering
  \includegraphics[scale=0.4]{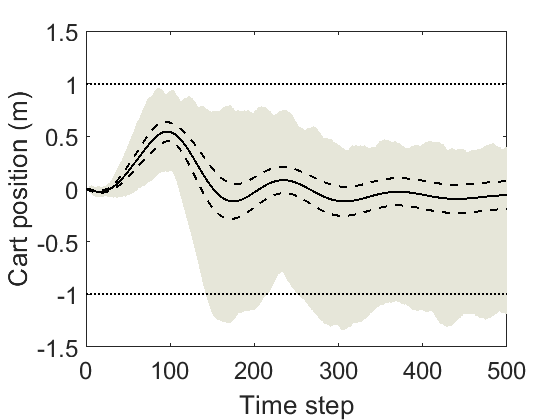}
  \centering
  \includegraphics[scale=0.4]{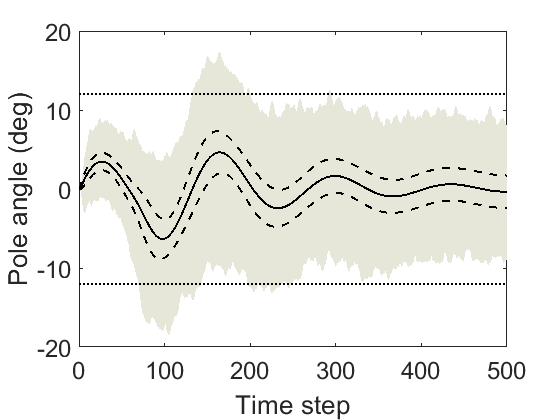}
\caption{\textbf{Cart-pole system.} Cart's position (left) and pole's angle (right) for $50,000$ trajectories under the learned policy. The grey area is an envelop for these trajectories, their mean is indicated by the solid line and the standard deviation around mean is indicated by dashed lines. Only $515$ trajectories ($1.03\%$) go outside of the safe location $[-1,1]$ or drop the pole outside of the angle interval $[-12^\circ,12^\circ]$.}
    \label{fig:sequential_learning}
\end{figure}

\begin{figure}
   \begin{minipage}{0.47\textwidth}
     \centering
     \includegraphics[width=1.05\linewidth]{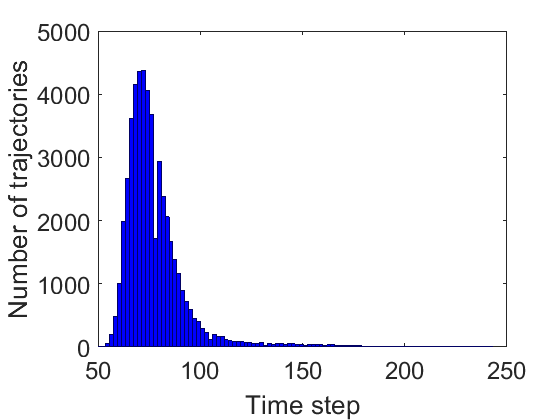}
     \caption{\textbf{Cart-pole system.} Histogram of the first time the trajectories reach the interval $[0.4,1]$. A majority of the trajectories reach this interval within $150$ time steps.}
     \label{Fig:hist}
   \end{minipage}\hfill
   \begin{minipage}{0.47\textwidth}
     \centering
     \includegraphics[width=1.05\linewidth]{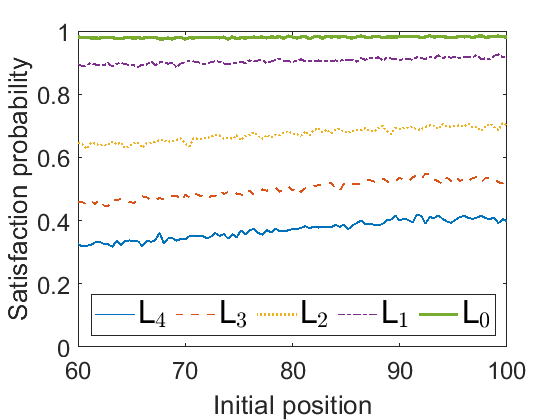}
     \caption{\textbf{Boat driving problem.} The satisfaction probability as a function of the initial position $y_0$ for the policies learned with labelling functions $\Lab_i$, $i\in\{0,1,2,3,4\}$.
     }
     \label{fig:sequential_learning_2}
   \end{minipage}
\end{figure}

Our sequential learning procedure successfully learns the policy within $44$ minutes and gives the lower bound $0.9526$ for satisfaction probability (according to Theorem~\ref{thm:augmented_lower_bound}). Figure~\ref{fig:sequential_learning} shows cart's position (left) and pole's angle (right) for $50,000$ trajectories under the learned policy. The grey area is an envelop for these trajectories, their mean is indicated by the solid line and the standard deviation around mean is indicated by dashed lines. Only $515$ trajectories ($1.03\%$) go outside of the safe location $[-1,1]$ or drop the pole outside of the angle interval $[-12^\circ,12^\circ]$. All trajectories reach the location $[0.4,1]$. The histogram of the first time the trajectories reach this interval is presented in Figure~\ref{Fig:hist}, which shows majority of the trajectories reach this interval within $150$ time steps.

Using Hoeffding's inequality,\footnote{Hoeffding's inequality asserts that the tail of the binomial distribution is exponentially decaying: $Prob(H\ge (p+\varepsilon)N)\le exp(-2\varepsilon^2 N)$ for all $\varepsilon>0$ with the number of trials $N$, the success probability $p$, and the observed number of successes $H$.} we get that the true satisfaction probability under the learned policy is in the interval $[0.975,1]$ with confidence $1-4\times 10^{-10}$. This is in line with the lower bound $0.9526$ computed by the RL.


\subsection{Boat Driving Problem}
\label{sec:boat}
The objective in the boat driving problem is to design a policy for driving a boat from the left bank to the right bank quay in a river with strong nonlinear current. Variations of this problem have been used in the literature (see e.g. \cite{279185}). We use a more general version presented in \cite{704563} with the dynamics reported in
the appendix (Section~\ref{sec:boat_dynamics}). 
The model has six continuous states including $x$ and $y$ coordinates for the location both in the interval $[0,200]$. The boat starts its journey from the left bank of the river $x_0=0$ and $y_0\in [60,100]$ and should reach the right bank of the river $x_n = 200$ and $y_n\in [95,105]$ for some $n$. There is an unknown nonlinear stochastic current affecting the location of the boat.

Direct application of A2C RL does not converge to a policy within $24$ hours. We then apply Alg.~\ref{al2} with labelling functions $\Lab_4, \Lab_3, \Lab_2, \Lab_1, \Lab_0$ respectively with the target range $[50,150]$, $[80,120]$, $[85,115]$, $[90,110]$, and $[95,105]$. We also adaptively increase the value of $\zeta$ to get better lower bounds on the satisfaction probability: 
$\zeta_4 = 0.9950$,
$\zeta_3 = 0.9965$,
$\zeta_2 = 0.9980$,
$\zeta_1 = 0.9995$, and
$\zeta_0 = 0.9999$.
%
%
The results of this sequential learning procedure are presented in Fig.~\ref{fig:sequential_learning_2} as a function of the initial position of the boat. The learning rate is set to $8\times 10^{-4}$ and the computational time is $70$ minutes. The results show that the lower bound on satisfaction probability is monotonically increasing for all initial positions of the boat when $\zeta$ increases, which shows also convergence as a function of $\zeta$.

In order to validate the computed bound, we took the initial position $(x_0,y_0) = (0,80)$ and obtained $50,\!000$ trajectories. All trajectories reach the target location. Based on Hoeffding's inequality, the true probability is in $[0.99,1]$ with confidence $5\times 10^{-5}$, which confirms the lower bound $0.9810$ computed by RL.






\section{Future Work}
We presented an approach for applying reinforcement learning (RL) to unknown continuous-state stochastic systems with the goal of satisfying a linear temporal logic specification. We formulated an optimal average reward criterion that converges linearly to the true optimal satisfaction probability under suitable assumptions. We used RL to learn a lower bound on this optimal value and improved the performance of the learning by a sequential algorithm using relaxed versions of the specification.
In future, we plan to study the relation with discounting reward functions \cite{bozkurt2019control}, formal connections with maximising frequency of visits,
and providing guidance in adapting the network architecture in the RL to the structure of the specification.


%
%
%
%
\bibliographystyle{splncs04}
\bibliography{main_arxiv_ifm}
\newpage
\section{Appendix}
\subsection{Proof of statements}
\label{sec:proofs}

\begin{proof}[Theorem~\ref{thm:augmented1}]
The mapping from $\bar\rho$ to $\rho$ is by restricting its domain to $\StS$: $\rho(\cdot|s) = \bar\rho(\cdot|s)$ for all $s\in\StS$. The mapping from $\rho$ to $\bar\rho$ is by extending the domain of $\rho$ with an arbitrary input choice at $\phi$:  $\bar\rho(\cdot|s) = \rho(\cdot|s)$ for all $s\in\StS$ and $\bar\rho(\cdot|\phi) = \boldsymbol{\delta}_{u}(\cdot)$ for some $u\in\IS$.\\
To prove the right-hand side of \eqref{eq:reach1}, we show that
$P_s^{\bar\rho}(\CMP_\zeta\models\square\neg \phi)\le 
    P_s^{\rho}(\CMP\models \lozenge \square \neg B).$
A path of $\CMP$ satisfying $\lozenge \square \neg B$ will take accepting states from $B$ only a finite number of times. We decompose the event $\CMP\models\lozenge \square \neg B$ into a set of disjoint events $\cup_{n=0}^\infty A_n$, where $A_n$ indicates the paths that visit the accepting states exactly $n$ times. Similarly, we decompose the event $\CMP_\zeta\models\square\neg \phi$ into $\cup_{n=0}^\infty B_n$, where $B_n$ indicates the paths of $\CMP_\zeta$ that visit the accepting states in $B$ exactly $n$ times and do not visit $\phi$. Then we have
\begin{eqnarray*}
    P_s^{\bar\rho}(\CMP_\zeta\models\square\neg \phi) & \myeq P_s^{\bar\rho}(\cup_{n=0}^\infty B_n) = \sum_{n=0}^\infty P_s^{\bar\rho}(B_n) =  \sum_{n=0}^\infty \zeta^n P_s^{\rho}(A_n)\\
    & \le \sum_{n=0}^\infty P_s^{\rho}(A_n)
      = P_s^{\rho}(\cup_{n=0}^\infty A_n)
     = P_s^{\rho}(\CMP\models \lozenge \square \neg B).
\end{eqnarray*}
Note that equality $(\ast)$ is true since the probability that paths of $\CMP_\zeta$ visit $B$ infinitely often and do not visit $\phi$ is zero. The left-hand side of \eqref{eq:reach1} is proved as follows:
\begin{eqnarray*}
    & P_s^{\rho}(\CMP\models \lozenge \square \neg B) - P_s^{\bar\rho}(\CMP_\zeta\models\square\neg \phi) = \sum_{n=0}^\infty P_s^{\rho}(A_n) - \sum_{n=0}^\infty \zeta^n P_s^{\rho}(A_n)\\
    & = \sum_{n=0}^\infty (1-\zeta^n)P_s^{\rho}(A_n)\le \sum_{n=0}^\infty n(1-\zeta) P_s^{\rho}(A_n)\\
 & = (1-\zeta) \sum_{n=0}^\infty n P_s^{\rho}(\tau_B=n)(1-P_s^{\rho}(\CMP\models \square\lozenge B)).
\end{eqnarray*}
The sum $\sum_{n=0}^\infty n P_s^{\rho}(\tau_B=n)$ is equal to $\mathbb E_s^\rho(\tau_B)$ which is less than or equal to $\tau^\ast_B$ and is a bounded quantity according to Assumption~\ref{ass:bounded}. Therefore,
\begin{eqnarray*}
    P_s^{\rho}(\CMP\models \lozenge \square \neg B&) - P_s^{\bar\rho}(\CMP_\zeta\models\square\neg \phi) \le (1-\zeta)\mathbb E_s^\rho(\tau_B)(1-P_s^{\rho}(\CMP\models \square\lozenge B))\\
    & \le (1-\zeta)\mathbb E_s^\rho(\tau_B).
\end{eqnarray*}
$\hfill\blacksquare$
\end{proof}

\begin{proof}[Theorem~\ref{thm:augmented_optimal}]
The proof of this theorem is by taking the supremum of the quantities in \eqref{eq:reach1} with respect to $\rho$ and $\bar\rho$ with a proper order:


\begin{align*}
    & P_s^{\rho}(\CMP\models \square\lozenge B) \le  P_s^{\bar\rho}(\CMP_\zeta\models\lozenge\phi)\le \sup_{\bar\rho}P_s^{\bar\rho}(\CMP_\zeta\models\lozenge\phi),\\
    & \Rightarrow
    P_s^{\rho}(\CMP\models \square\lozenge B) \le \sup_{\bar\rho}P_s^{\bar\rho}(\CMP_\zeta\models\lozenge\phi),\quad\forall\rho.
\end{align*}
The first inequality is due to \eqref{eq:reach1} and the second one is due to the definition of supremum. We now take supremum of both sides with respect to $\rho$ to get
\begin{align*}
    \sup_{\rho} P_s^{\rho}(\CMP\models \square\lozenge B) \le \sup_{\bar\rho}P_s^{\bar\rho}(\CMP_\zeta\models\lozenge\phi).
\end{align*}
The second inequality is proved similarly. Note that
\begin{align*}
P_s^{\bar\rho}(\CMP_\zeta\models\lozenge\phi) & \le
     P_s^{\rho}(\CMP\models \square\lozenge B) + (1-\zeta)\mathbb E_s^\rho(\tau_B)\\
    & \le  \sup_{\rho}P_s^{\rho}(\CMP\models \square\lozenge B) + (1-\zeta)\sup_{\rho}\mathbb E_s^\rho(\tau_B)\\
    \Rightarrow
    P_s^{\bar\rho}(\CMP_\zeta\models\lozenge\phi) & \le \sup_{\rho}P_s^{\rho}(\CMP\models \square\lozenge B) + (1-\zeta)\tau^\ast_B,\quad\forall\bar\rho.
\end{align*}
The first inequality is due to \eqref{eq:reach1} and the second one is due to the definition of supremum and Assumption~\ref{ass:bounded} that ensures $\tau^\ast_B$ is bounded. We now take supremum of both sides with respect to $\bar\rho$ to get
\begin{equation*}
    \sup_{\bar\rho}P_s^{\bar\rho}(\CMP_\zeta\models\lozenge\phi) \le \sup_{\rho}P_s^{\rho}(\CMP\models \square\lozenge B) + (1-\zeta)\tau^\ast_B.
\end{equation*}
$\hfill\blacksquare$
\end{proof}

\begin{proof}[Lemma~\ref{lem:policy}]
The LDBA $\mathcal A$ and any positional policy $\Cont$ can be used to construct a finite-memory policy $\rho_f = (\hatS,\hat s_0,\Kerp,\Kero)$ for $\CMP$ as defined in Definition~\ref{def:memory_policy}. $\hatS$ will be the state space of $\mathcal A$, $\hat s_0$ its initial state, $\Kerp$ its transition function combined with the labelling function, and $\Kero$ the positional policy $\Cont$.  

Due to the first part of the lemma, to show that finite-memory policies are sufficient for solving Problem~\ref{prob:LTL}, it is sufficient to show that positional policies are sufficient for solving problem~\ref{prob:Buchi}. 
The B\"uchi condition $\psi_B$ is a special case of two-player zero-sum stochastic games. More specifically,
\begin{equation}
\label{eq:game}
 P_s^{\rho}(\CMP\models \square\lozenge B) = P_s^{\rho}\left(\limsup_n{\mathbf{1}_B(s_n)}=1\right) = \mathbb E_s^{\rho}\left[\limsup_n{\mathbf{1}_B(s_n)}\right].  
\end{equation}
Therefore, solving Problem~\ref{prob:Buchi} is equivalent to solving a two-player game with limsup payoff \cite{MS03}, where one player is maximising \eqref{eq:game} by choosing a policy and the other player has a singleton choice. This class of games is studied in \cite{FPS_games_18} for continuous state spaces and it is proved that if the problem admits an optimal policy, it has a positional policy.~\hfill$\blacksquare$
\end{proof}

\begin{proof}[Theorem~\ref{thm:augmented_lower_bound}]
The second inequality is immediate from Theorem~\ref{thm:augmented_optimal}, the product construction, and Lemma~\ref{lem:policy}. For the first inequality, define $\CMP^\otimes_{1}$ as the product of $\CMP$ and $\LDBA_{\neg\psi}$ with accepting states $\Acc_1^\otimes$. Then, 
\begin{align*}
P_s^{\rho}(\CMP\models \psi) = 1- P_s^{\rho}(\CMP\models\neg\psi) & \myeq 1- P_{s,q_0}^{\rho}(\CMP^\otimes_{1}\models\always\eventually \Acc_1^\otimes)\\
& \mygeqq 
    1-P_{s,q_0}^{\bar\rho}(\CMP^\otimes_{1\zeta}\models\lozenge\phi),
\end{align*}
where $(\ast)$ follows from the product construction and $(\ast\ast)$ from Theorem~\ref{thm:augmented_optimal}. Taking supremum w.r.t. $\rho$ and then w.r.t. $\bar\rho$ concludes the proof of \eqref{eq:lower_bound}.
\hfill$\blacksquare$
\end{proof}

\begin{proof}[Theorem~\ref{thm:relaxed}]
The claim follows inductively on the structure of $\psi$ in its PNF. All the operators in PNF are monotonic with respect to satisfaction of their sub-formulas. Therefore, if \eqref{eq:sequence_r} holds for $\psi_1$ and $\psi_2$, it is also true for $\psi_1\vee\psi_2$, $\psi_1 \wedge \psi_2$, $\nex \psi_1$ $\psi_1\until \psi_2$, and $\psi_1 \release \psi_2$. The claim is also true on atomic propositions and their negations due to Theorem~\ref{thm:sequence_r}.
$\hfill\blacksquare$
\end{proof}

\subsection{Pseudo Algorithm for A2C Learning on the Augmented CMP}
\label{sec:A2C}
A pseudo algorithm of our approach based on the A2C is provided in Alg.~\ref{al1}. The inputs of the algorithm are $\CMP$ as a black box that generates the data, the specification $\psi$, and the labelling function $\Lab$.
The algorithm outputs two neural networks $\mu(s,q|\theta^{\mu})$ and $\mathcal Q(s,q|\theta^{\mathcal Q})$ for the actor and for the critic, respectively. $\mu(s,q|\theta^{\mu})$ is the policy on the product space and $\mathcal Q(s,q_0|\theta^{\mathcal Q})$ is the learned probability values as a function of initial state of $\CMP$.
The hyper-parameters of the algorithm are $\zeta<1$, $\textsf{episode\_number}$, $\textsf{episode\_horizon}$, and neural network architectures for Actor and Critic.
In each learning episode,
data is gathered from the CMP under the policy $\mu$ (Steps~\ref{step:init},\ref{step:input},\ref{step:state}). The associated paths of the LDBA $\LDBA_\psi$ is also computed (Step~\ref{step:mode}) and the total reward is obtained depending on the samples from a Bernoulli random variable modelling the effect of $\zeta$ (Steps~\ref{step:random_zeta},\ref{step:reward_zeta}). This information is used to update the parameters of the Actor and Critic networks for next episode (Step~\ref{step:update}).  

Note that the set of valid inputs $\IS_\zeta(s,q)$ is state dependent. We encode this in the Actor network by penalising the invalid inputs and freezing the state of the CMP if such inputs are taken. We also use a technique called \emph{hot-encoding} \cite{GoodBengCour16,HARRIS2016108} to make the neural networks suitable for combination of both discrete $(q)$ and continuous $(s)$ values.

\begin{algorithm2e}[t]
\label{al1}
\DontPrintSemicolon
\SetKwInOut{Input}{input}
\SetKwInOut{Output}{output}
\SetKw{return}{return}
\SetKwRepeat{Do}{do}{while}
\SetKwData{conflict}{conflict}
\SetKwFor{Loop}{Loop}{}{}
\SetKw{KwNot}{not}
\caption{Maximising $P_s^{\bar\rho}(\CMP^\otimes_\zeta\!\models\!\lozenge\!\phi)$ using Advantage Actor-Critic RL}
\begin{small}
    \Input{CMP $\CMP$ as a black box, LTL specification $\psi$, labelling function $\Lab$}
    \Output{Actor network $\mu(s,q|\theta^{\mu})$ and Critic network $\mathcal Q(s,q|\theta^{\mathcal Q})$}
    Select \textbf{hyper-parameters}: $\zeta<1$, $\textsf{episode\_number}$, $\textsf{episode\_horizon}$, neural network architectures for Actor and Critic\\
    Convert $\psi$ to an LDBA $\mathcal A_\psi = (Q,\Sigma,\delta,q_{0},\Acc)$ as in Section~\ref{subsec:LDBA}\;
    Set $A^\epsilon := \{\epsilon_q|q\in Q\}$ and $\Acc^\otimes$ according to \eqref{eq:accepting}\;
    Initialise $\theta^{\mu}$ and $\theta^{\mathcal Q}$ randomly\;
    \For{$\textsf{t=0}$ \textbf{to} $\textsf{episode\_number}$}
    {   
         Pick initial state $s_0^\otimes=(s_0,q_0)$ randomly and set $R(s_0,q_0) = 0$\label{step:init}\;
        \Repeat{$X_\zeta=1$ or end of episode}
        {   
            Choose input $u_n$ by sampling from $\mu(s_n,q_n|\theta^{\mu})$\label{step:input}\\
            \If{$u_n\in A^\epsilon$}
            { $s_{n+1} = s_n$ and $q_{n+1} = q'$ with any $u_n = \epsilon_{q'}$}
            \If{$u_n\in \IS$}
            {Apply input $u_n$ to $\CMP$ and observe the next state $s_{t+1}$\label{step:state}\;
            Set $q_{n+1} = \delta(q_n,\Lab(s_n))$\label{step:mode}\;
            }
            \If{$(s_n,q_n)\in \Acc^\otimes$}{Set $X_\zeta$ to a value in $\{0,1\}$ randomly with $Prob(X_\zeta=0) = \zeta$\label{step:random_zeta}\;
            Set $R(s_n,q_n)=1$ if $X_\zeta=1$\label{step:reward_zeta}}
            %
        }
        Store paths $(s_0,q_0,u_0,s_1,q_1,u_1,\ldots)$ and their rewards $\sum_n\! R(s_n,q_n)$\;
            Update parameters $\theta^{\mathcal Q}$ and $\theta^\mu$ by minimising the Actor and Critic loss functions using observed paths and their rewards\label{step:update}

    }
\end{small}
\end{algorithm2e}

\subsection{Dynamics in the Boat Driving Problem}
\label{sec:boat_dynamics}
The new position of the boat are
\begin{equation}
    \begin{array}{l}
    x_{n+1}= \min(200, \max (0, x_n+v_{n+1}\cos(\delta_{n+1})))\\
    y_{n+1}= \min(200, \max (0, y_n-v_{n+1} \sin(\delta_{n+1})-E(x_{n+1},\eta_{n+1}))).
    \end{array}
\end{equation}
The boat angle $\delta_n$ and velocity $v_n$ are computed as
\begin{equation}
    \begin{array}{l}
    \delta_{n+1} = \delta_{n} + I\Omega_{n+1}\\
    \Omega_{n+1} = \Omega_{n} + (\omega_{n+1} - \Omega_n)(v_{n+1}/v_{\textsf{max}})\\
    v_{n+1} = v_n + I(v_{\textsf{desired}}-v_n)\\
    \omega_{n+1} = \min(\max(p(u_{n}-\delta_{n}),-45^\circ),45^\circ),
    \end{array}
\end{equation}
where $\Omega_n$ is the angular velocity,
$\omega_n$ is the rudder angle,
$I$ is the system inertia,
$v_{\textsf{max}}$ is the maximum speed,
$v_{\textsf{desired}}$ is the desired speed of the boat,
and $p$ is the proportional gain that relates the desired direction $u_n$ to the rudder angle.
The input $u_n$ can take $12$ directions ranging from southwest to north
\begin{equation*}
    \IS= \{-100^\circ, -90^\circ, -75^\circ, -60^\circ, -45^\circ, -30^\circ, -15^\circ, 0^\circ, 15^\circ, 45^\circ, 75^\circ, 90^\circ\}.
\end{equation*}
the current effect is a nonlinear function of horizontal distance from the left bank $x$. It is defined by $E(x,\eta)= f_c[\frac{x}{50}-(\frac{x}{100})^2]+\eta$ where $f_c$ is the current force and $\eta$ is the disturbance. We use $f_c= 1.25$, $I=0.1$, $v_m= 2.5$, $v_d=1.75$, and $p=0.9$ for generating data from the system and assume $\eta$ has normal distribution. Note that we treat the model as black box and use it to generate data. The dynamics are not known to our learning algorithms.

\end{document}